\documentclass[preprint,groupedaddress,superscriptaddress,amsmath,amssymb]{revtex4-1}[10pts]
\usepackage{amsmath}
\usepackage{amssymb}
\usepackage{float}
\usepackage[pdftex]{graphicx}
\usepackage{color}
\usepackage{graphicx}
\usepackage{dcolumn}
\usepackage{bm}
\usepackage{color}
\usepackage{comment}
\usepackage{soul}
\usepackage{ulem}
\usepackage[noabbrev]{cleveref}
\graphicspath{{figures/}}

\begin{document}
\title{Tunable X-band optoelectronic oscillators based on external-cavity semiconductor lasers}
\author{C.~Y.~Chang}
\email{cychang@gatech.edu} 
\affiliation{Georgia Institute of Technology, School of Physics, Atlanta, Georgia 30332-0250 USA}%
\affiliation{%
UMI 2958 Georgia Tech-CNRS, Georgia Tech Lorraine, 2 Rue Marconi F-57070, Metz, France}%

\author{Michael J. Wishon}
\affiliation{%
	UMI 2958 Georgia Tech-CNRS, Georgia Tech Lorraine, 2 Rue Marconi F-57070, Metz, France}%
\affiliation{School of Electrical and Computer Engineering, 
Georgia Institute of Technology, Atlanta, Georgia 30332-0250 USA}%

\author{Daeyoung Choi}%
\affiliation{%
	UMI 2958 Georgia Tech-CNRS, Georgia Tech Lorraine, 2 Rue Marconi F-57070, Metz, France}%
\affiliation{School of Electrical and Computer Engineering, 
Georgia Institute of Technology, Atlanta, Georgia 30332-0250 USA}%

\author{K. Merghem}
\author{Abderrahim Ramdane}
\affiliation{%
Center for Nanosciences and Nanotechnologies (CNRS-C2N), Route de Nozay, 91460 Marcoussis-France}%

\author{Fran\c{c}ois Lelarge}
\email{Present address: Almae Technologies, Route de Nozay, 91460 Marcoussis- France}
\affiliation{%
Center for Nanosciences and Nanotechnologies (CNRS-C2N), Route de Nozay, 91460 Marcoussis-France}%


\author{A. Martinez}
\affiliation{%
Center for Nanosciences and Nanotechnologies (CNRS-C2N), Route de Nozay, 91460 Marcoussis-France}%

\author{A. Locquet}
\affiliation{%
	UMI 2958 Georgia Tech-CNRS, Georgia Tech Lorraine, 2 Rue Marconi F-57070, Metz, France}%
\affiliation{School of Electrical and Computer Engineering, Georgia Institute of Technology, 
Atlanta, Georgia 30332-0250 USA}%

\author{D. S. Citrin}
\email{david.citrin@ece.gatech.edu}
\affiliation{%
	UMI 2958 Georgia Tech-CNRS, Georgia Tech Lorraine, 2 Rue Marconi F-57070, Metz, France}%
\affiliation{School of Electrical and Computer Engineering, 
Georgia Institute of Technology, Atlanta, Georgia 30332-0250 USA}%

\date{\today}
\begin{abstract}
Laser diodes with optical feedback can exhibit periodic intensity oscillations at or near the 
relaxation-oscillation frequency.  We demonstrate 
optoelectronic oscillators based on external-cavity semiconductor 
lasers in a periodic dynamical regime tunable over the entire X-band. 
Moreover, unlike standard optoelectronic oscillators, we need not employ 
the time-dependent optical intensity incident on a photodiode to generate the microwave signal, 
but rather have the option of generating the electrical microwave 
signal directly as a voltage $V(t)$ at the laser-diode injection terminals under constant current operation; 
no photodiode need be involved, thus circumventing optical-to-electrical conversion. 
We achieve a timing jitter of $\lesssim 10$ ps and a quality factor of $\gtrsim 2\times 10^5$ 
across the entire X-band, that ranges from 6.79 GHz to 11.48 GHz. Tuning is achieved by varying the injection current $J$.
\end{abstract}
\maketitle
\section{Introduction}
Microwave optoelectronic oscillators (OEO) have attracted attention due to the tunability and stability of 
low-power laser diodes (LD) \citep{williamson,maleki}. OEOs enable tremendous flexibility; the optical signal can be converted 
immediately to a microwave electrical signal via a fast photodiode (PD) or can be transmitted over 
low-loss optical fiber to be converted downstream to a microwave electrical signal, again by a PD.  
 
There are several approaches to achieving OEOs. One common approach is to beat two 
phase-locked optical waves \citep{yao1,yao2}; others are based on optical injection of a master laser 
into a slave laser \citep{15Nelson} or electro-optic modulators \citep{96Yao,10Chembo}.

While in some implementations, the optical signal is used to convey the microwave 
information for later optical-to-electrical (O/E) conversion, for a number of applications, 
the optical signal [time-dependent optical intensity $I(t)$] itself is of no intrinsic interest.
Instead, O/E conversion is carried out proximate to the optical generation by one or more PDs.
Nonetheless, in all cases of which we are aware, a PD separate from the LD is required to 
generate the electrical signal.  Eliminating the O/E conversion, therefore, would be a 
simplification in a large class of OEO-based systems.

\begin{figure}[b]
\vspace{-.6cm}
\centerline{\includegraphics[width=0.6\textwidth]{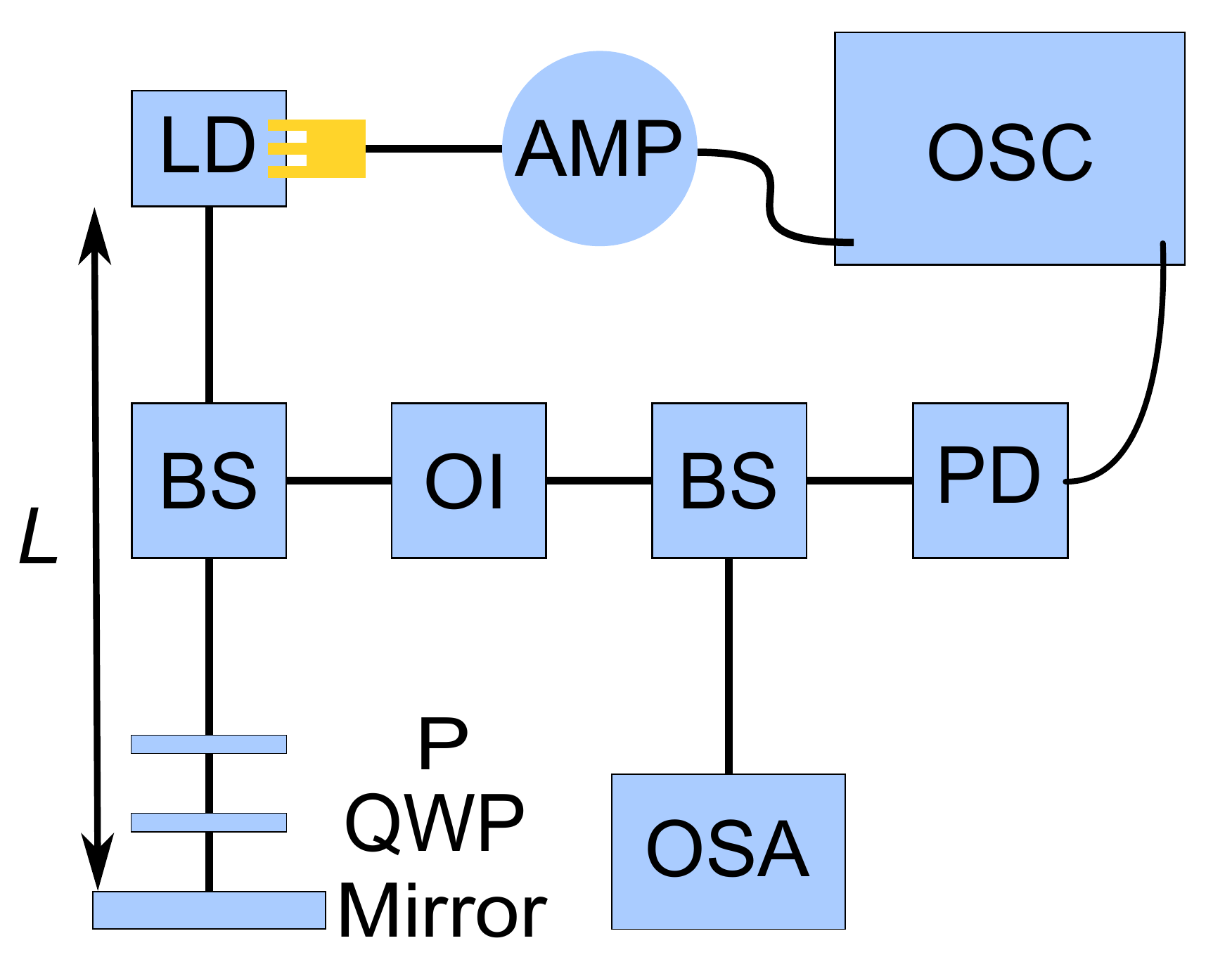}}
\caption{Experimental setup. LD: laser diode, AMP: RF amplifier, OSC: oscilloscope, BS: beam splitter, 
OI: optical isolator, PD: photodiode, P: polarizer, QWP: quarter-wave plate, and OSA: optical spectrum analyzer.
}
\label{figure1}
\end{figure}

We point the way to do so. This work demonstrates an OEO based on an external-cavity 
semiconductor laser (ECL) in which the microwave electrical signal is generated directly
by monitoring the voltage $V(t)$ across the injection terminals of the LD under 
constant-current $J$ conditions; in addition, the periodic optical intensity $I(t)$ may also be used for microwave generation if so desired. An approach extracting electrical microwave signals in an OEO has been demonstrated in Ref. \citep{IEEEPTLZhao} in a structure incorporating an electroabsorption modulated laser (EML). We point out that in our case, no EML is required; the direct optical feedback onto the LD produces spontaneous intensity oscillations.
In broad terms, in a periodic dynamical regime in an ECL, oscillations in $I(t)$ and $V(t)$ 
occur that are $\sim\pi$ out of phase (though the true dynamics as predicted by the Lang-Kobayashi equations \citep{lk}
are somewhat different),
\textit{i.e.}, one observes undamped relaxation oscillations at frequency $f_{RO}$. With the monotonic dependency of $f_{RO}$ on the injection current, the tunability across the entire X-band is achieved by varying the injection current $J$.
As is well known, $V(t)$ is directly related to the inversion $N(t)$ in the LD
active region  \citep{74kazarinov,14sahai,06Roy}. We find, for our LD, that the amplitude of the oscillations in $V(t)$ 
is around 278 $\mu V$, the oscillation frequency is tunable from 6.79 to 11.48 GHz 
(the upper limit here is due to the frequency cutoff of our oscilloscope). The oscillation frequency is created by a Hopf bifurcation of an external-cavity mode. It has been shown that its frequency close to the relaxation-oscillation frequency depends on the pumping current, $J$ and the feedback strength \citep{YarivOxford, CohenJQE,AcketJQE}. 
Typical values of the timing jitter are $\lesssim 10$ ps, and the quality factor $Q$ is  
$\gtrsim 2 \times 10^5$.  The combination of wide tunability and low noise figures of merit 
may make this device competitive with state-of-the art OEOs.

\section{Experiment}

A schematic diagram of the experiment is illustrated in Fig.\ 1. 
The ECL is based on a single longitudinal-mode edge-emitting InGaAsP DFB LD containing 
seven quantum-wells in the active region. The grating is designed and fabricated to achieve a 
$k$ factor of 50 cm$^{-1}$ and the length $l$ of the LD is measured to be 0.6 mm, 
resulting in a $kl$ value of 3. The LD  emits at 1550 nm with free-running threshold current $J_{th}=29.8$ mA. 
The structure has been described 
in detail and investigated for feedback tolerance in Ref. \citep{AnthonyAPL}. 
For the ECL, the experimental feedback strength $\eta$ is 
determined by the relative angle between a polarizer (P) and a quarter-wave plate (QWP) in Fig. 1. The maximum feedback strength $\eta = 1$ corresponds to $\sim$16 \% of the optical power being coupled back onto the collimating lens.
The QWP is mounted on a motorized rotational stage with a step size of $0.01^{\circ}$. 
For the measurement of $V(t)$, a RF probe (Cascade Microtech AE-ACP40-GSG-400) with 40-GHz 
bandwidth is employed to extract $V(t)$ from the LD injection terminals. 
The AC and DC components of $V(t)$ are separated with a bias tee (Keysight 11612A), 
and  amplified with an 18-dB amplifier (Newport 1422-LF) with 20 GHz bandwidth. 
In addition, the AC component of $I(t)$ and $V(t)$ are simultaneously  
recorded on a real-time oscilloscope (OSC) (Agilent DSO80804B)
with  12 GHz cut-off frequency. An optical spectrum analyzer (BOSA 400) is used to detect the 
purity of the optical signal. The external cavity length $L$ is chosen to be 42, 68, or 94~cm, 
corresponding to an external cavity round-trip time of $\tau=$ 2.8, 4.5, or 5.9 ns and giving an external-cavity 
free-spectral range of $f_{\tau} = \tau^{-1}=$ 0.36, 0.22, or 0.17 GHz.  

\begin{figure}[b]
\vspace{-.3cm}
\centerline{
\includegraphics[width=0.5\textwidth]{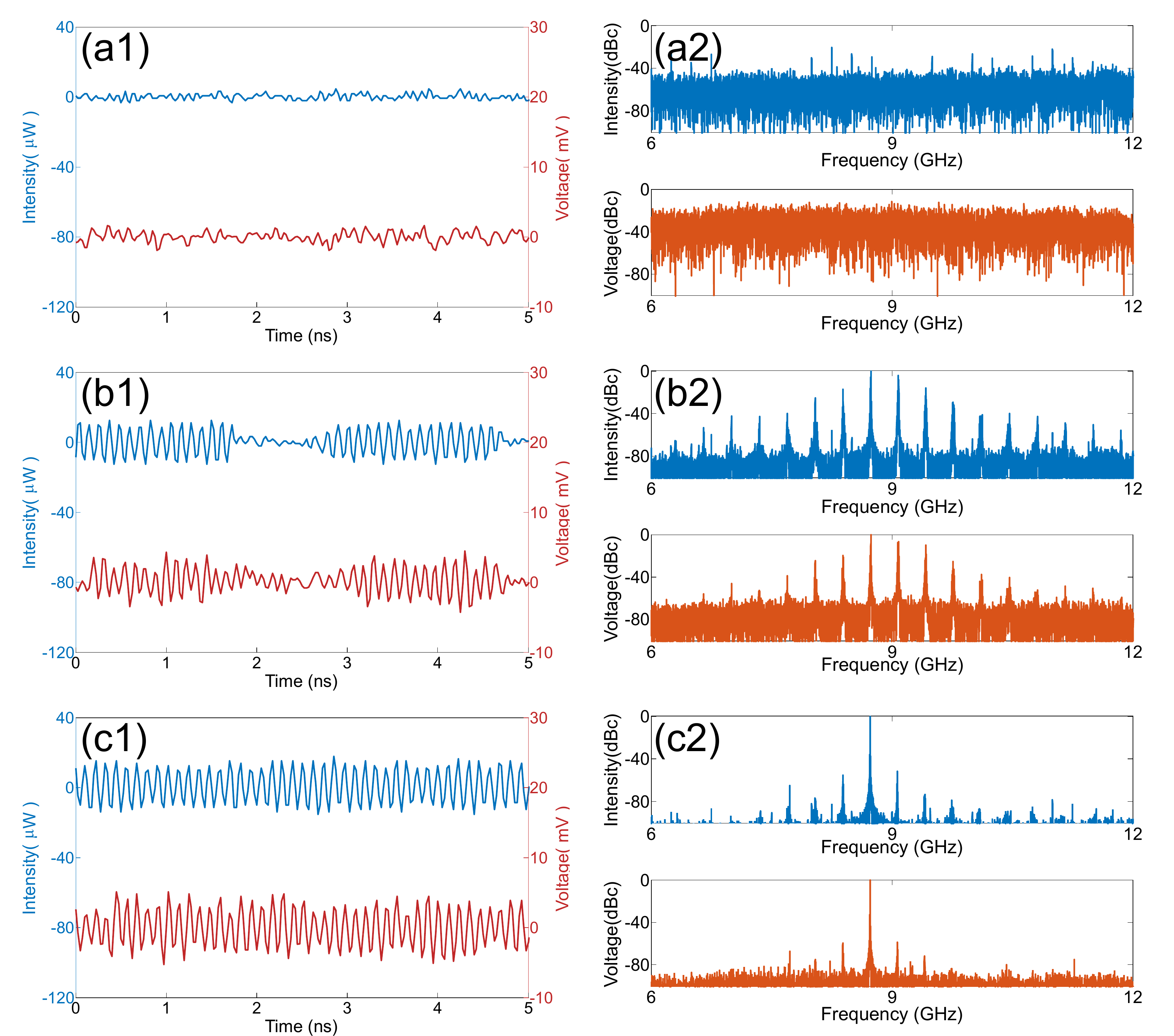}
}
\caption{The left panels show $I(t)$ 
and $V(t)$ in blue and orange, respectively, for  (a) $\eta=0.01$, 
(b) $0.19$, and (c) $0.28$ for $L = 42 $ cm and $J = 70$ mA. 
The right panels are the RF spectra of the time series to the left;  (a) 
CW, (b) quasi-periodicity, and (c)  periodicity.
}
\label{figure2}
\end{figure}

\section{Results and Discussion}

\begin{figure}[bt]
\vspace{-.4cm}
\centerline{\includegraphics[width=0.5\textwidth]{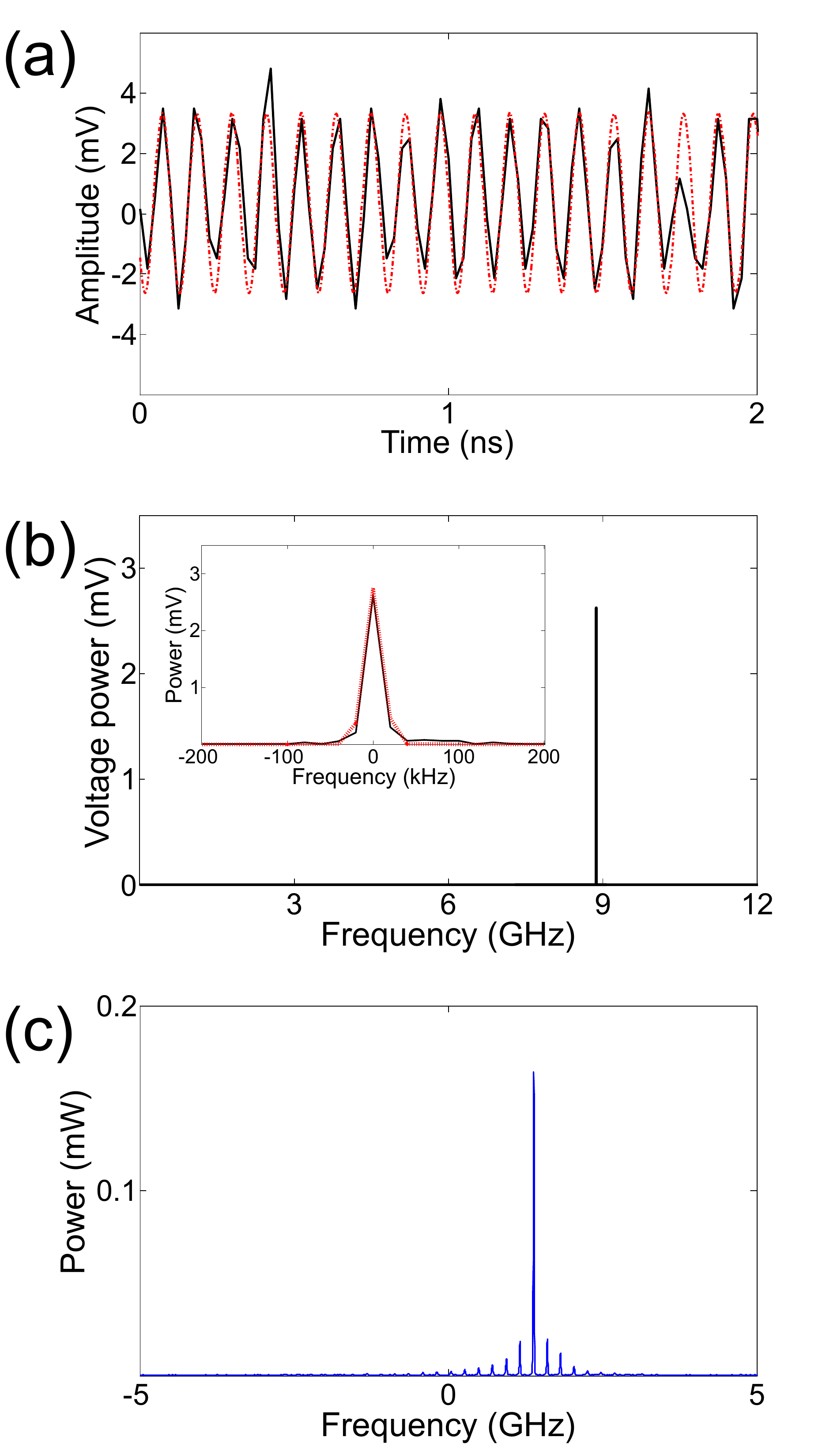}}
\vspace{-.4cm}

\caption{(a) $V(t)$ (black) and the fitted 
sinusoid (red dotted line). (b) RF spectrum of $V(t)$ 
while the inset shows the spectrum on an expanded frequency scale (adjusted so that 
0 kHz lies at line center) ($Q \approx 3.5 \times 10^5$). A Gaussian fit of the spectrum is represented with a red dotted line. 
(c) Corresponding optical spectrum, where the frequency of the solitary laser is set to zero. All data are acquired simultaneously with
$J = 70$ mA, $L = 68$ cm, and $\eta=0.19$.
}
\label{figure3}
\vspace{-.5cm}
\end{figure}

\begin{figure}[ht]
\vspace{-.5cm}
\includegraphics[width=0.48\textwidth]{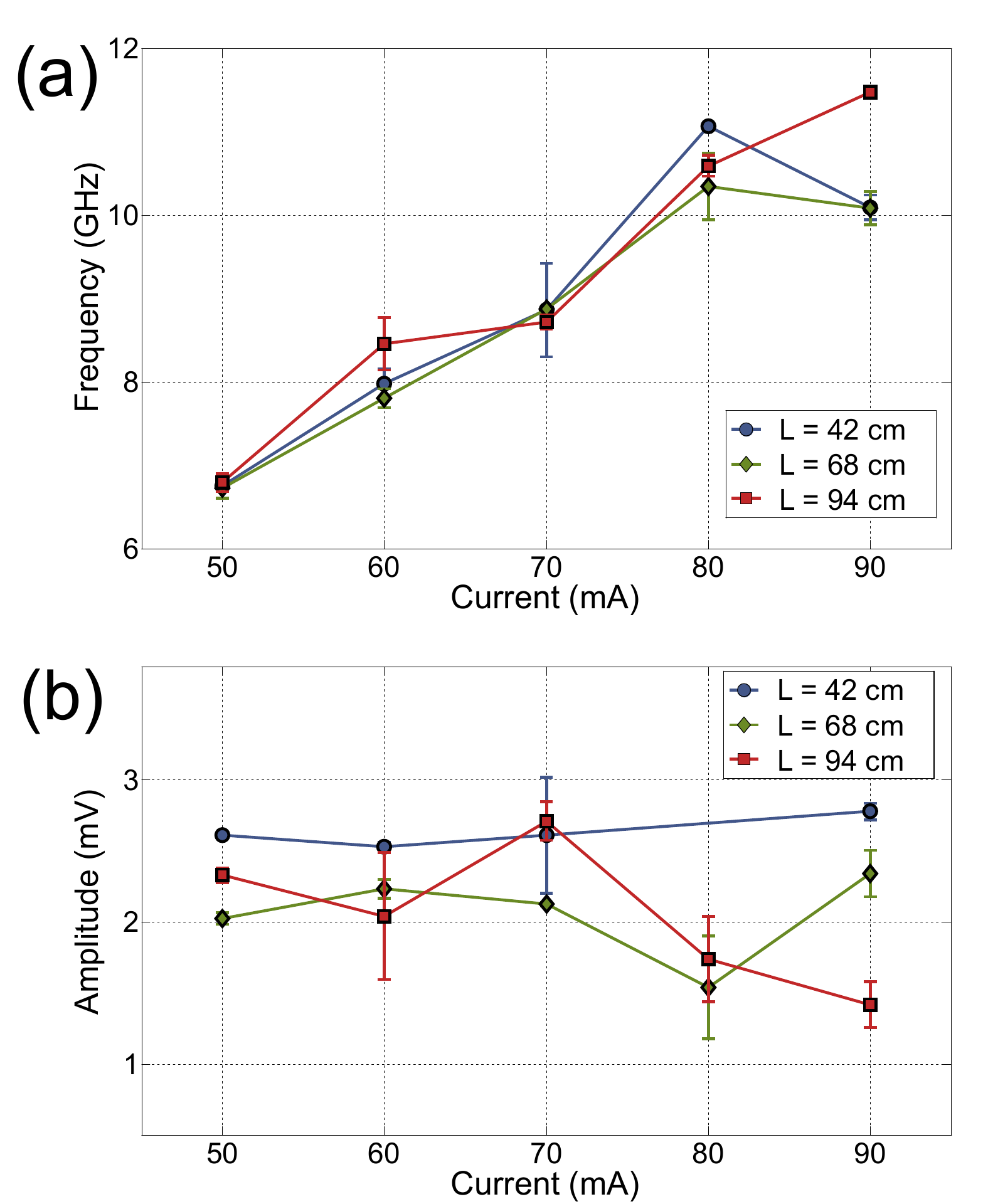}
\caption{(a) Measured frequencies and (b) amplitudes 
as functions of $J$ at $L= 42$, 68, and 94 cm with $\eta=$ 0.19. }
\label{figure4}
\vspace{-.6cm}
\end{figure}

The various dynamical regimes (including chaos) accessed by our ECL are amply
discussed in Ref. \citep{CYCAPL} to which we refer the interested reader.
Our aim here is to focus on the regime in which $I(t)$ and $V(t)$ are periodic.
To illustrate the progression from CW operation to periodic oscillations, in Fig.\ 2 are 
shown simultaneous time series for $I(t)$ and $V(t)$ with the corresponding RF 
spectra for various $\eta$ at  $J = 70$ mA $\approx 2.3 J_{th}$ and $L=42$ cm ($f_{\tau}=0.36$ GHz),
where $f_{RO}=8.87$ GHz. Moving down Fig.\ 2, $\eta$ increases:
$\eta=0.01,0.19, 0.28$.  For very low $\eta=0.01$, the ECL output is similar to that of the solitary LD showing
CW behavior. Here, both $I(t)$ and $V(t)$ are relatively constant apart from noise originating in part from
spontaneous emission. As $\eta$ is increased to 0.19, the ECL has entered a quasi-periodic regime.  The dynamics
are dominated by the beating $f_{RO}=8.87$ GHz with $f_{\tau}=0.36$ GHz as seen in Fig.\ 2(b1). 
The RF spectra of Fig.\ 2(b2) for $I(t)$ and $V(t)$ are both peaked at $f_{RO}$ with sidebands 
separated from the maximum by multiples of $f_{\tau}$. For larger $\eta=0.19$, a periodic dynamical regime is 
finally accessed  [Fig.\ 2(c1)]; this is the dynamical regime of interest to the present study.  
Here both $I(t)$ and $V(t)$ are highly periodic; the RF spectra 
[Fig.\ 2(c2)] have narrowed considerably as compared with Fig.\ 2(b2).  Figure 2(c2) shows that the 
spectrum of the periodic signal is characterized by a dominant main peak at $f_{RO}=8.87$ GHz 
and sidebands separated by $f_{\tau}$. Also, $I(t)$ and $V(t)$ are $\sim\pi$ out of phase reflecting the interchange of optical and material excitation due to relaxation oscillations.
\begin{figure}[ht]
\vspace{-.3cm}
\includegraphics[width=0.48\textwidth]{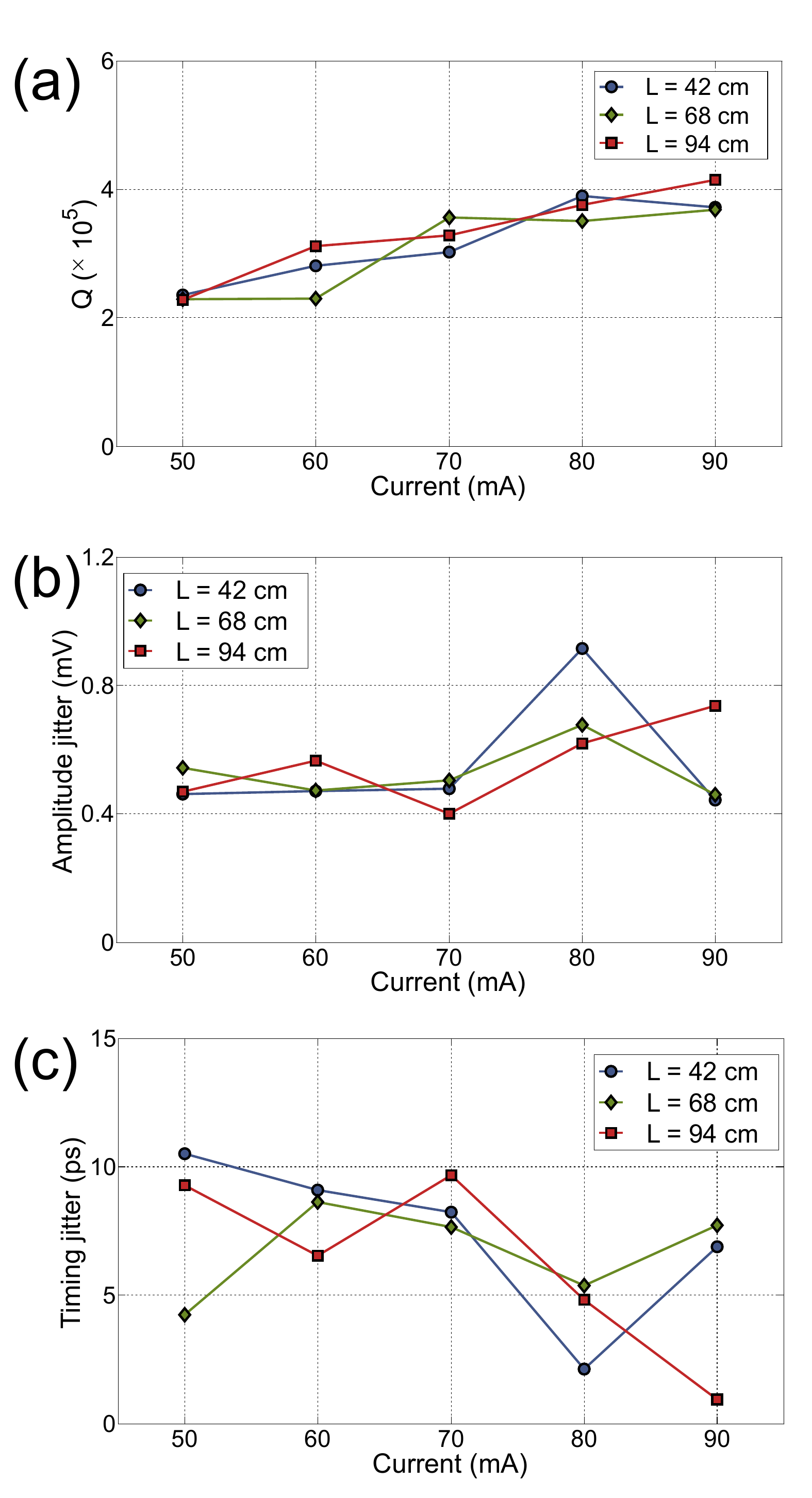}

\caption{(a) Quality factor $Q$, (b) amplitude jitter $\sigma_{amp}$, and (c) timing jitter
$\sigma_{t}$ of the OEO as a function of $J$  at $L= 42$, 68, and 94 cm with $\eta=0.19.$  
 }
\label{figure5}
\vspace{-.5cm}
\end{figure}

We now focus on the case when $J=70$ mA, where the relaxation-oscillation frequency 
$f_{RO}=8.87$ GHz, $L = 68$ cm ($f_{\tau}=0.22$ GHz), and $\eta=0.13$.  
In addition, we heretofore concentrate on $V(t)$, though similar dynamics are exhibited in $I(t)$. 
A typical time series for $V(t)$ is shown in Fig.\ 3(a)
$J = 70$ mA, $L = 68$ cm, and $\eta=$ 0.19.  Also shown is a sinusoid 
fit to the measured time series.
The signal is seen to be highly periodic, 
though contains both amplitude and timing jitter (see below for quantitative characterization).
Note that this signal has been amplified with a gain 18 dB, which may increase jitter.
The RF spectrum [Fig.\ 3(b)] is narrow and centered on $f_{RO}=8.87$ GHz with weak sidebands separated 
by $f_{\tau}$. In Fig. 3(c), we present the optical spectrum of $I(t)$ and observe that one 
external cavity mode (ECM) dominates. This is the 6th ECM mode to the right of the minimum linewidth mode\citep{bobbyOE,bobbyPRA}. Its frequency is of 1.39 GHz on the optical-frequency scale 
of Fig. 3(c) corresponding to an ECM at 1552.07 nm (193.33 THz).

We have carried out similar measurements for $J=50$, $60$, $70$, $80$, and $90$ mA and $L= 42$, 68, and 94 cm.  
Figure 4(a) shows the center frequency as a function of $J$ for 
$L= 42$, 68, and 94 cm with $\eta=0.19$.
We see that with ECLs 
based on this single LD, tunability between 6.79 to 11.48 GHz is 
achievable. Thanks to the strong variation of the relaxation-oscillation frequency of a laser diode with the injection current \citep{YarivOxford}. The upper frequency cutoff, however, corresponds to the bandwidth of our oscilloscope; 
independent measurements of $f_{RO}$ for this LD \citep{AnthonyAPL}
show $f_{RO}$ as high as 13 GHz.
Moreover, Fig.\ 4(b) shows that after amplification (18 dB), electrical signals with few-millivolt rms 
peak value are consistently obtained.

\begin{table*}
\centering
\vspace{.12cm}
\caption{A comparison of state of the art OEOs. All phase noise values are at 10 kHz offset.}
\begin{tabular}{|c|c|c|c|c|}
\hline
 System                & Fabry-Perot     & Bragg Grating  & Optical Filter   & Our System    
 \\ \hline 
 Phase Noise (dBc/Hz)  & $-92.8$         & $-102$         & $-120$           
             & $-80$ (estimated)           
             \\
 Tunability (GHz)      & $6.41 - 10.85$  & $3 - 28$       & $4.74 - 38.38$   & $6.79 - 11.48$   
\\\hline
\end{tabular}
\vspace{-.12cm}
\end{table*}

To evaluate the performance of OEOs, it is important to evaluate
the quality factor $Q$, the amplitude jitter $\sigma _{amp}$, and the timing jitter $\sigma _{t}$. $Q$ 
is defined as the ratio of an oscillator's carrier frequency to the full width at half maximum in its power spectrum and it can be used to estimate the bit-error rate of a digital communication 
system when the decision variable is assumed to be Gaussian \citep{Agrawal,BerganoPTL,MateraQE}. 
Here, $Q$ is determined by fitting the RF spectrum to a Gaussian curve, as shown in the inset of 
Fig 3(b). In Fig. 5(a), $Q$ 
as a function of $J$  at $L= 42$, $68$, and $94$ cm with $\eta=0.19$ 
is determined to be $\gtrsim 2 \times 10^5$, which is 
confirmed with a RF spectrum analyzer (Anritsu MS2830A).
Amplitude jitter $\sigma _{amp}$ is
calculated by 
demodulating with a sinusoid at the center frequency of $V(t)$ and removing the resulting high-frequency 
term. 
Typical values of amplitude jitter $\sigma_{amp}$ seen in 
Fig.\ 5(b) are between 0.4 and 0.9 mV and, as a result, reduce $Q$. 
Timing jitter $\sigma _{t}$ is also determined by demodulation and ranges from $0.9$ to $10.52$ ps. Side peaks separated, approximately, by multiples of $f_{\tau}$ can be observed in the RF spectrum of Fig. 2(c). We determined that the peak-to-pedestal ratio, based on the largest side peak, is typically larger than 25 dB.

\begin{figure}[ht]

\vspace{-.5cm}
\includegraphics[width=0.5\textwidth]{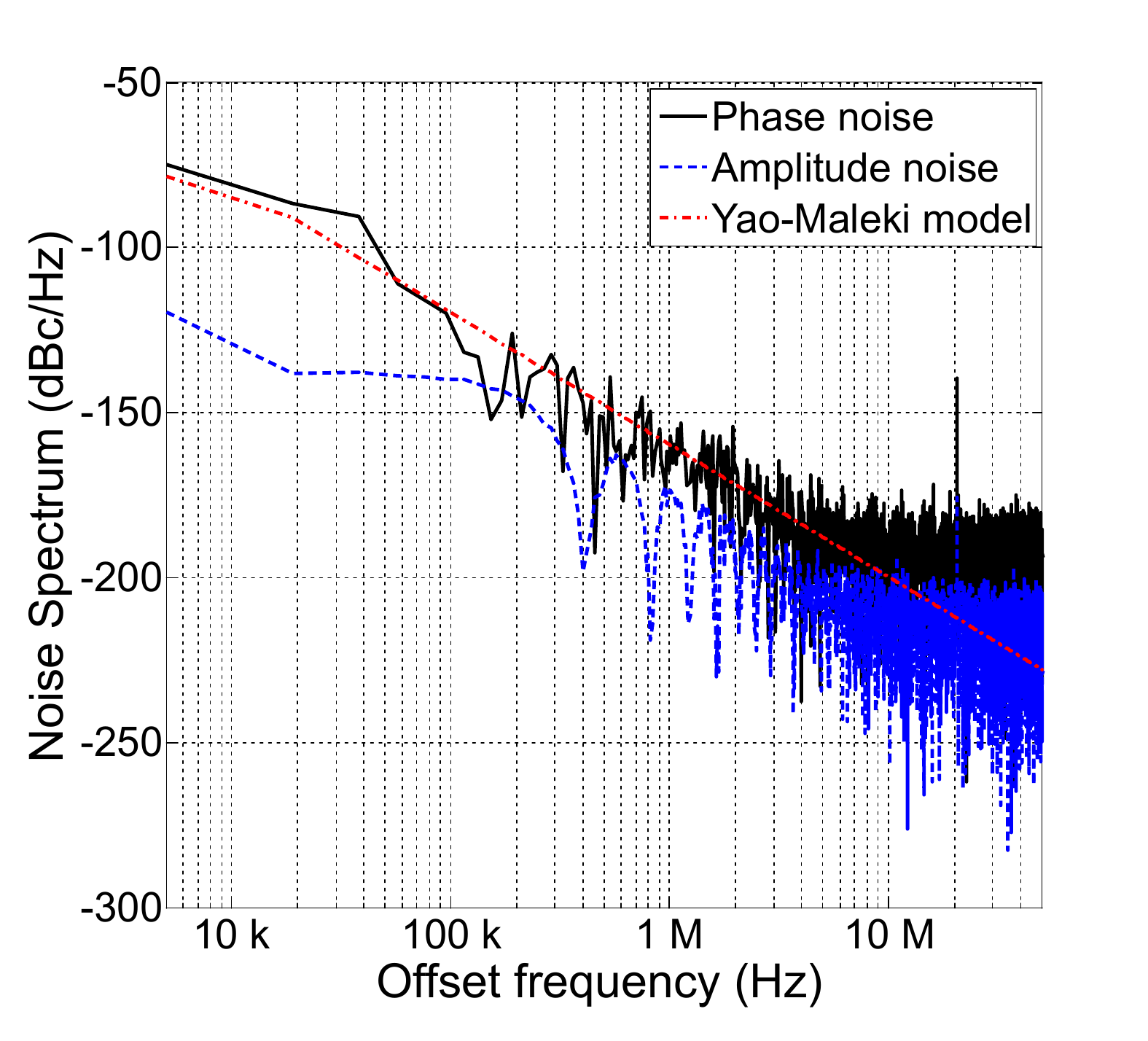}
\vspace{-.6cm}
\caption{The noise spectrum for amplitude (blue) and phase noises (black) with $J = 90 $mA, $L = 94$ cm, and
$\eta=0.19$. 
The phase noise is determined to be $-79.3$ dBc/Hz at 10 kHz offset frequency by fitting with a Yao-Maleki model\citep{96Yao}.}
\label{figure6}
\vspace{-.4cm}
\end{figure}

In Fig. 6, we plot the noise density spectrum for both amplitude and phase
for $J=90$ mA, $L=94$ cm, and $\eta= 0.19$. We determined the phase noise to be -86.3 dBc/Hz at 20 kHz offset frequency. 
Due to resolution limitations, we are unable to obtain a direct measurement of the phase-noise spectral density
at 10 kHz offset; however, we have fit our phase-noise spectrum to a Yao-Maleki model \citep{Yao:96}
and extrapolated the phase noise to be $-79.3$ dBc/Hz at 10 kHz offset frequency. In addition, we confirmed that the value is smaller than -80 dBc/Hz for all values of current and delay tested in this paper. 

Our OEO compared to tunable state of the art X-band OEOs can be seen in Table I. 
The OEOs used for comparison are an OEO based on a laser whose output is modulated 
by an external modulator before being optically injected into a Fabry-Perot LD \citep{YaoFabry}, 
an OEO using two cascaded phase modulators followed by a 
linearly chirped fiber Bragg grating \citep{YaoBragg}, and an OEO which 
uses a phase modulator followed by a tunable optical filter \citep{ChenOEFilter}. 
The tunability of our OEO ($6.79$ to $11.48$ GHz) is slightly lower than two of the 
reported OEOs (the upper limit results from our oscilloscope bandwidth).
Choosing a different LD, however, may result in a larger range of accessible $f_{RO}$.
 The phase noise of our OEO is 
somewhat larger than the competing approaches. 
However, it must be pointed out that our oscillator has a high $Q$ factor and the 
advantage of a simpler design by not requiring optical to electronic conversion as in the other OEOs.
Nor has our OEO been actively stabilized, and our noise measurements have been obtained following amplification which is likely to contribute
to those measurements.

\section{Conclusion}
We have demonstrated the use of an ECL as a novel OEO tunable across the entire X-band, 
from 6.79 to 11.48 GHz. Both the optical $I(t)$ and electrical $V(t)$ signal can be 
employed for microwave applications; however, by using $V(t)$ directly (upon which we concentrate here), 
we can entirely eliminate the need for O/E conversion,
which is of interest for some applications. Specifically, 
$V(t)$ is obtained by monitoring the voltage across the LD injection terminals 
under constant-current operation. The quality factor $Q$ is greater than $2 \times 10^5$. In addition, 
we have characterized both amplitude and timing jitters
with $\sigma_{amp}$ from 0.40 mV to 0.92 mV and $\sigma_t \lesssim 10.52$ ps. 
An estimated value of the phase noise spectral density at 10 kHz offset is found to be $-80$ dBc/Hz. The phase-noise performance of our OEO setup may be improved by choosing the feedback time to be a multiple of the relaxation-oscillation time; in this way, one may average out phase fluctuations \citep{KathyAPL,LinaPRA}. Alternatively, a Pyragas-like feedback can be applied to control unwanted chaos and stabilize unstable orbits \citep{PyragasPRA}. This type of feedback is achieved by applying a continuous feedback term which is proportional to the difference of signal at $t$ and $t - \tau$ to the system. Such feedback scheme can be implemented by modulating the injection current $I(t)$ with optoelectronic feedback as shown in \citep{Turovets} or with impulsive delayed feedback \citep{Naumenko}. The combination of tunability and low phase-noise figures of merit indicates that the OEOs based on
ECLs may be competitive with the state of the art. Future efforts will be geared towards stabilizing the device and determining the intrinsic performance before amplification.
\section*{Acknowledgment}

We gratefully acknowledge the financial support of the R\'{e}gion Grand Est.


\end{document}